# Pressure-induced reentrant superconductivity in a misfit layered compound $(SnS)_{1.15}(TaS_2)$

Chutong Zhang,[1,2,*] Jiajia Feng,[3,*] Xiao Tang,[1,2,*] Xiangzhuo Xing,[1,2,†] Na Zuo,[1,2] Xiaolei Yi,[4] Yan Meng,[5] Xiaoran Zhang,[1,2,‡] Rajesh Kumar Ulaganathan,[6] Raman Sankar,[7] Xiaofeng Xu,[8] Xin Chen,[1,2,§] and Xiaobing Liu[1,2]

[1]*Key Laboratory of Quantum Materials under Extreme Conditions in Shandong Province, School of Physics and Physical Engineering, Qufu Normal University, Qufu 273165, China*

[2]*Laboratory of High Pressure Physics and Material Science (HPPMS), Advanced Research Institute of Multidisciplinary Sciences, Qufu Normal University, Qufu 273165, China*

[3]*Center for High Pressure Science and Technology Advanced Research, Beijing 100193, China*

[4] *College of Physics and Electronic Engineering, Xinyang Normal University, Xinyang 464000, China*

[5]*School of Physics and Electronic Engineering, Jining University, Qufu 273155, China*

[6]*Centre for Nanotechnology, Indian Institute of Technology Roorkee, Roorkee 247667, India*

[7]*Institute of Physics, Academia Sinica, Taipei 11529, Taiwan*

[8]*School of Physics, Zhejiang University of Technology, Hangzhou 310023, China*

[*]These authors contributed equally to this work.

Corresponding authors: [†] xzxing@qfnu.edu.cn, [‡] xiaoran_zhang@qfnu.edu.cn, [§] chenxin@qfnu.edu.cn

## Abstract

Misfit layered compounds are natural van der Waals heterostructures in which electronically active transition-metal dichalcogenide layers are decoupled by incommensurate blocking layers, enabling bulk realization of quasi-two-dimensional quantum states. Here we investigate the superconducting, transport, and structural properties of the misfit compound $(SnS)_{1.15}(TaS_2)$ under pressures up to ~150 GPa. The low-pressure superconducting phase is gradually suppressed and disappears near 14.7 GPa, accompanied by increasing residual resistance. Remarkably, a distinct superconducting phase reemerges above ~80 GPa and persists to the highest pressures achieved. This reentrant superconductivity follows a pressure-induced sign reversal of the Hall coefficient near 60 GPa and a nonmonotonic evolution of the normal-state resistance, indicating an electronic reconstruction. No structural phase transition is detected over the entire pressure range. Our results demonstrate a pressure-driven electronic reconstruction leading to reentrant superconductivity in a misfit layered compound, establishing pressure as an effective route to engineer superconductivity and electronic states in natural van der Waals heterostructures.

Two-dimensional (2D) van der Waals heterostructures (vdWHs) have emerged as a versatile platform for exploring novel quantum phenomena over the past decade [1-3], driven by rapid advances in the isolation, assembly, and manipulation of atomically thin materials [4-7]. By vertically stacking dissimilar 2D layers with clean and weakly bonded interfaces, vdWHs not only integrate the intrinsic functionalities of their constituent materials but also give rise to emergent properties through interlayer proximity effects. Such engineered heterostructures have enabled the realization of a wealth of exotic phenomena[1-3, 8-11], highlighting the power of interfacial design in quantum materials.

Misfit layered compounds represent a unique and naturally occurring class of vdWHs, in which alternating rock-salt-type monochalcogenide blocking layers and transition-metal dichalcogenide (TMDC) active layers are stacked along the out-of-plane direction with incommensurate lattice periodicities in-plane [12]. These compounds are generally described by the formula $(MX)_{1+\delta}(TX_2)$ ($M$ = Sn, Pb, Bi, rare earth; $T$=transition metal; $X$ = S, Se), where the misfit parameter $\delta$ reflects the lattice mismatch between the constituent layers. Originally synthesized in the 1970s and structurally characterized in the 1980s [13, 14], misfit compounds have recently attracted renewed interest [12, 15-27]. The insertion of the monochalcogenide blocking layers effectively weakens or even eliminates interlayer coupling between adjacent TMDC sheets, while simultaneously protecting the active layers from environmental degradation [18, 20]. As a result, misfit compounds offer an exceptional opportunity to access monolayer-like electronic properties within a stable bulk crystal, preserving the 2D nature even in the bulk form [18-21, 26]. Analogous to intercalation-induced effects, substantial interlayer charge transfer has been observed in these systems [20, 24, 28], playing a key role in stabilizing the complex superlattice structures. Moreover, the reduced dimensionality combined with strong spin-orbit coupling in the decoupled TMDC layers has been shown to generate a variety of emergent phenomena, including Ising and unconventional superconductivity [17, 21-23, 27], diverse magnetic orders [26], topologically driven anomalous transport [15, 25], and, more recently, the coexistence of robust in-plane ferroelectricity and high metallicity in weakly coupled misfit systems [16].

Pressure has proven to be a powerful tuning parameter for modulating crystal structure and electronic properties in both vdWH and bulk TMDC systems [29-39], by continuously reducing the interlayer spacing. Despite extensive high-pressure studies on layered materials, the pressure response of misfit layered compounds remains largely unexplored. Given their intrinsic heterostructured nature and highly tunable interlayer interactions, pressure provides an ideal avenue to manipulate interlayer coupling, charge transfer, and electronic states in misfit systems. In this study, we performed high-pressure studies on a representative misfit compound $(SnS)_{1.15}(TaS_2)$ by employing diamond anvil cell (DAC) technique. Upon compression, the low-pressure superconducting state is progressively suppressed and vanishes at approximately 14.7 GPa. Remarkably, with further increasing pressure, a distinct high-pressure superconducting phase reemerges near 80 GPa. High-pressure x-ray diffraction (XRD) measurements reveal that, aside from the expected lattice compression, no change in crystal symmetry occurs over the entire pressure range studied. Moreover, the reentrant superconductivity emerges following a pressure-induced sign reversal of the Hall coefficient near 60 GPa, signalling an electronic reconstruction that could be beneficial for the reappearance of superconductivity. These results reveal a close interplay between interlayer coupling, electronic structure reconstruction, and emergent superconductivity in misfit layered compounds.

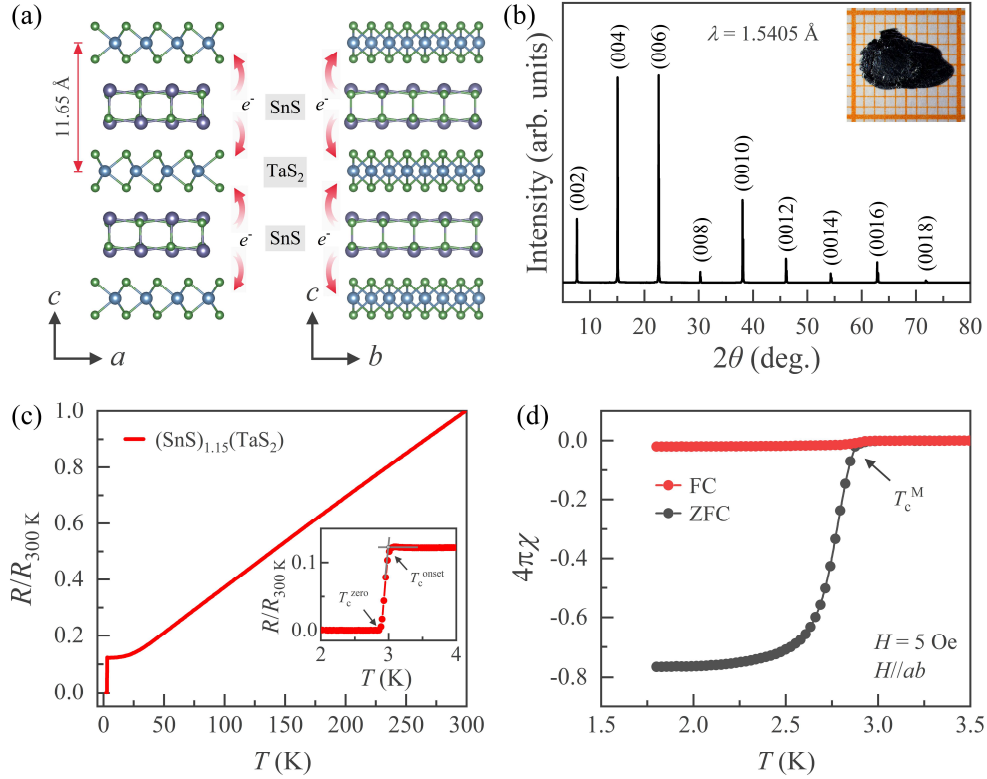

FIG. 1. Structure and basic physical characterization of misfit compound $(SnS)_{1.15}(TaS_2)$. (a) Schematic crystal structure of $(SnS)_{1.15}(TaS_2)$ viewed along different crystallographic directions. Red arrows indicate interlayer charge transfer from SnS layers to $TaS_2$ layers. (b) Single-crystal XRD pattern collected using Cu $K\alpha$ radiation. Inset: optical photograph of a representative single crystal. (c) Normalized temperature-dependent resistance at ambient pressure. Inset: enlarged view highlighting the superconducting transition. $T_c^{onset}$ is defined as the crossover point between the normal-state resistance and the superconducting transition. (d) Superconducting diamagnetic response measured under a magnetic field of 5 Oe with $H//ab$, shown for both zero-field-cooled (ZFC) and field-cooled (FC) conditions.

The experimental methods used in this study, as well as some supporting materials, are given in the Supplemental Material [40] (see also references therein [41, 42]). The $(SnS)_{1.15}(TaS_2)$ superlattice consists of alternating $1H$-$TaS_2$ layers and rock-salt-type SnS blocking layers, as schematically illustrated in Fig.1 (a). Figure 1 (b) displays the singe-crystal XRD pattern of synthesized $(SnS)_{1.15}(TaS_2)$, in which only (00$l$) diffraction peaks are observed, indicating excellent $c$-axis orientation. Figure 1(c) shows the normalized temperature-dependent resistance, revealing metallic behavior over the entire measured temperature range and a sharp superconducting transition at low temperatures. The onset superconducting transition temperature $T_c^{onset}$ is approximately 3.0 K, while the zero-resistance temperature $T_c^{zero}$ is about 2.9 K. Superconductivity is further corroborated by magnetization measurements [Fig. 1(d)], which exhibit a clear diamagnetic transition at $T_c^M \sim 2.9$ K, consistent with that of resistance measurement. Notably, the superconducting transition temperature of $(SnS)_{1.15}(TaS_2)$ is significantly enhanced compared to that of bulk $2H$-$TaS_2$ ($\sim 0.8$ K) [43] and closely matches that reported for monolayer $TaS_2$ [44, 45], indicating that the SnS blocking layers effectively decouple the individual $1H$-$TaS_2$ sheets and stabilize monolayer-like superconductivity in the bulk superlattice.

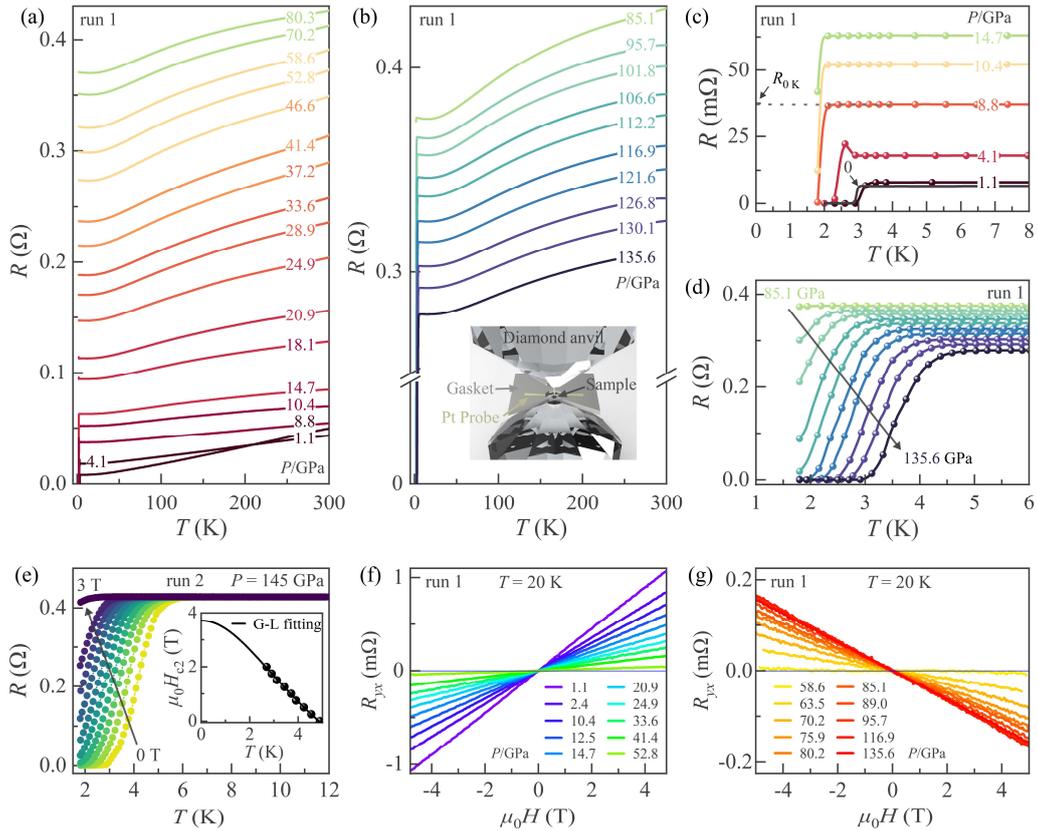

FIG. 2. Electrical transport properties of $(SnS)_{1.15}(TaS_2)$ under pressure. (a)-(b) Temperature-dependent resistance ($R$-$T$) under various pressures in run 1. Inset in (b) shows the schematic of the DAC configuration. (c)-(d) Enlarged views of the low-temperature $R$-$T$ curves near the superconducting transition. The $R$-$T$ curve at $P = 0$ GPa from Fig. 1(c), with a rescaled vertical axis, is included for comparison. The dashed line in (c) indicates a linear extrapolation of the normal-state resistance, which is used to determine the residual resistance $R_{0\,K}$. (e) $R$-$T$ curves near the superconducting transition measured under various magnetic fields at 145 GPa in run 2. Inset: upper critical fields $\mu_0 H_{c2}$ as a function of temperature. The solid line represents a fitting using the G-L model. (f)-(g) Hall resistance as a function of magnetic field measured at 20 K under various pressures in run 1.

Figure 2 presents the electrical transport properties of $(SnS)_{1.15}(TaS_2)$ under pressure in run 1. It is seen that the normal-state resistance remains metallic behavior at all measured pressures [Figs. 2(a) and 2(b)]. With increasing pressure, the overall resistance first increases, reaching a maximum near 80.3

GPa [Fig. 2(a)], and then decreases upon further compression [Fig. 2(b)]. Superconductivity is slightly enhanced at 1.1 GPa compared to ambient pressure [Fig. 1(c)], but is subsequently suppressed with increasing pressure and vanishes above 14.7 GPa. Unexpectedly, at higher pressures, a subtle resistance drop reappears at low temperatures starting from 85.1 GPa [Fig. 2(d)], which becomes more pronounced and ultimately evolves into a zero-resistance state, signaling the reentrant superconductivity. Similar behavior was reproduced in a second independent measurement (run 2), as shown in Fig. S1 [40], confirming the reproducibility of the observed phenomena. Figure 2(e) presents the temperature-dependent resistance measured under various magnetic fields near the superconducting transition at 145 GPa. The superconducting transition systematically shifts to lower temperatures with increasing magnetic field. The inset of Fig. 2(e) shows the temperature dependence of the upper critical field, $\mu_0 H_{c2}(T)$, determined using the criterion of 90% of the normal-state resistance. The data are well described by the Ginzburg-Landau (G-L) model, $\mu_0 H_{c2}(T) = \mu_0 H_{c2}(0)[1-(T/T_c)^2]/[1+(T/T_c)^2]$, yielding $\mu_0 H_{c2}(0) \approx 3.8$ T. To further elucidate the evolution of the electronic structure, we measured the field-dependent Hall resistance $R_{yx}(H)$ at 20 K under various pressures, as shown in Figs. 2(f) and 2(g). In all cases, $R_{yx}(H)$ varies linearly with magnetic field. Notably, the slope of $R_{yx}(H)$ decreases monotonically with pressure and changes sign from positive to negative above 58.6 GPa, revealing a pressure-induced crossover of the dominant charge carriers from hole-type to electron-type.

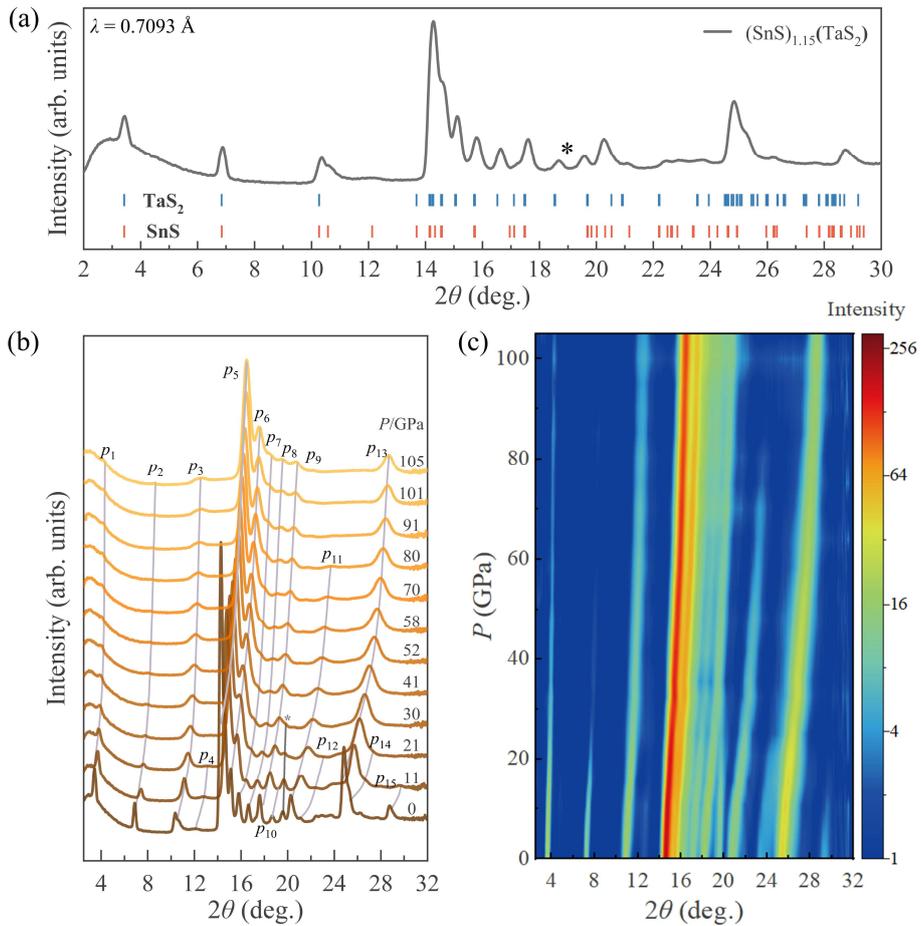

FIG. 3. High-pressure XRD measurements. (a) Powder XRD pattern collected at ambient pressure using Mo $K\alpha$ radiation. The peak marked by an asterisk originates from the $c$-BN coating on the Re gasket. (b) Evolution of the XRD patterns with increasing pressure. The vertical curves are guides to the eye, illustrating the evolution of the diffraction peaks with pressure. (c) Contour plot of the XRD patterns under pressure.

To elucidate the electronic evolution and examine whether the reentrant superconductivity is associated with a structural phase transition, we performed *in-situ* high-pressure XRD measurements. Figure 3(a) shows the powder XRD pattern of $(SnS)_{1.15}(TaS_2)$ at ambient pressure, where diffraction peaks from both the SnS and $TaS_2$ sublattices are clearly resolved [46]. Figures 3(b) and 3(c) present the pressure evolution of the XRD patterns in linear and contour representations, respectively. Owing to the large number of reflections from the two alternating sublattices, the merging of peaks, and additional peak broadening induced by pressure inhomogeneity at high pressures, precise indexing of individual reflections becomes unreliable. We therefore label the dominant diffraction peaks generically as $P_n$ ($n$ = 1, 2, 3, …). The corresponding pressure dependence of the $d$-spacings is summarized in Fig. S2 [40]. Upon compression, all diffraction peaks systematically shift toward higher angles, reflecting lattice contraction, while the overall diffraction pattern remains unchanged compared to that at ambient pressure. Although the peak intensities gradually weaken and the number of observable reflections decreases at high pressures, primarily due to the limited sample volume and geometric constraints of the diamond anvil cell, no new peaks appear throughout the measured pressure range. Upon decompression to ambient pressure, all diffraction peaks observed at ambient conditions are fully recovered (see Fig. S3 [40]), demonstrating the reversibility of the structural response. These results indicate the absence of a pressure-induced structural transition, thereby excluding structural symmetry changes as the origin of the anomalous electronic evolution and reentrant superconductivity in $(SnS)_{1.15}(TaS_2)$.

We summarize the onset superconducting transition temperature $T_c^{onset}$, the resistance at 300 K ($R_{300 K}$), the residual resistance at 0 K ($R_{0 K}$), and the Hall slope ($\alpha = R_{yx}/H$) as a function of pressure and construct the high-pressure phase diagram, as shown in Fig. 4. In the low-pressure superconducting phase (SC-I), $T_c^{onset}$ initially increases slightly upon compression, consistent with our previous measurements using a commercial piston-cylinder pressure cell [41] (see Fig. S4 [40]). With further increasing pressure, $T_c^{onset}$ is gradually suppressed and completely vanishes above 14.7 GPa. Upon further compression, a second superconducting phase (SC-II) emerges at approximately 80 GPa. In this high-pressure regime, superconductivity is continuously enhanced, persisting up to the pressure of ~150 GPa, the maximum pressure achieved in this work. Meanwhile, both the values of $R_{300 K}$ and $R_{0 K}$ exhibit pronounced nonmonotonic, dome-like pressure dependences, and the Hall slope undergoes a sign reversal with increasing pressure.

Notably, suppression of $T_c$ in the SC-I phase has also been reported in bulk 2$H$-$TaS_2$, although its superconductivity survives to much higher pressures and with a higher $T_c$ [32, 33]. In 2$H$-$TaS_2$, this behavior has been attributed to enhanced impurity scattering, as evidenced by the increase in residual resistivity [32]. According to Anderson's theorem, conventional superconductivity is robust against weak disorder introduced by nonmagnetic impurities [47]; however, sufficiently strong disorder can suppress superconductivity [48-50]. In $(SnS)_{1.15}(TaS_2)$, the residual resistance $R_{0\,K}$, which reflects impurity scattering, also increases steadily with pressure up to approximately 80 GPa. This trend suggests that the suppression of superconductivity in the SC-I phase is closely linked to enhanced impurity scattering. As a misfit compound, $(SnS)_{1.15}(TaS_2)$ intrinsically hosts significant structural distortions arising from lattice mismatch between the SnS and $TaS_2$ sublayers, which can strongly influence its physical properties [12]. External pressure may further amplify these distortions, thereby increasing disorder-induced scattering.

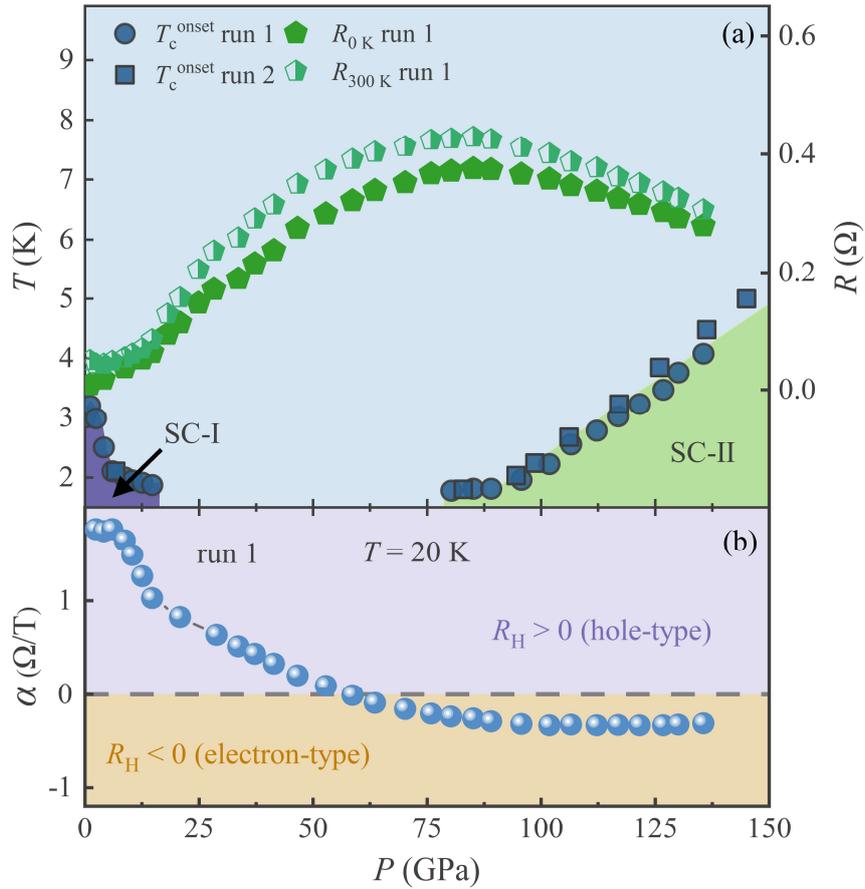

FIG. 4. High-pressure phase diagram of $(SnS)_{1.15}(TaS_2)$. Pressure dependence of (a) onset superconducting transition temperature $T_c^{onset}$, normal-state resistance at 300 K, residual resistance, and (b) Hall slope at 20 K with pressure. "SC" in (a) indicates the superconducting phase. Horizontal dashed line in (b) is included as visual guides.

For the high-pressure SC-II phase, our high-pressure XRD measurements have ruled out a pressure-induced structural phase transition as the origin of reentrant superconductivity, pointing instead to an electronic mechanism. First, the emergence of SC-II coincides with a pronounced reduction in the normal-state resistance, indicating a close connection between enhanced charge transport and superconductivity. Second, the dominant carrier in the SC-II region is electron-type, in stark contrast to the hole-dominated transport in the SC-I phase. The sign reversal of the Hall slope near 60 GPa provides evidence for an electronic structure reconstruction, which can be regarded as a precursor to the emergence of SC-II. Under pressure, the reduced interlayer spacing (see Fig. S2 [40]) enhances interlayer coupling mediated by charge transfer from the SnS sublayers to the $TaS_2$ sublayers, leading to a reconstruction of the tailored electronic structure of $(SnS)_{1.15}(TaS_2)$ [16, 36]. In addition, pressure may modify the degree of lattice mismatch or the interlayer stacking angle between alternating SnS and $TaS_2$ sublayers due to the weak van der Waals interlayer interactions. Such effects can induce charge redistribution between adjacent heterolayers, offering further opportunities for pressure-driven electronic structure reconstruction. A similar pressure-induced band-structure modification mechanism has been proposed in restacked $TaS_2$ [34], where random stacking of multiple $2H$-$TaS_2$ monolayers leads to tunable electronic properties through pressure-controlled interlayer stacking angles. A similar mechanism is likely operative in $(SnS)_{1.15}(TaS_2)$, providing a natural explanation for the reentrant superconductivity observed at high pressures. Notably, the superconducting transition temperature of the high-pressure SC-II phase in $(SnS)_{1.15}(TaS_2)$ remains significantly lower than that of bulk $2H$-$TaS_2$ and restacked $TaS_2$ [33, 34]. This difference is plausibly attributed to the presence of the inserted SnS blocking layers, which weaken the effective interlayer coupling compared with the more strongly coupled $TaS_2$ systems.

In conclusion, we have investigated the pressure-dependent superconducting, transport, and structural properties of the misfit layered compound $(SnS)_{1.15}(TaS_2)$ up to ~150 GPa. The low-pressure superconducting phase is progressively suppressed and vanishes near 14.7 GPa. At higher pressures, superconductivity reemerges above ~80 GPa and persists to the highest pressures achieved. This reentrant superconductivity is preceded by a sign reversal of the Hall coefficient and a nonmonotonic evolution of the normal-state resistance, signaling an electronic structure reconstruction. High-pressure XRD measurements reveal no structural phase transition, establishing an electronic origin of the reentrant superconducting phase. Our results demonstrate that pressure provides an effective route to engineer electronic structure and superconductivity in misfit layered compounds.

We thank Dr. Zejun Li for helpful discussions. This work was supported by the National Natural Science Foundation of China (Grant Nos. 12574020, 12474017, 12204265, and 12274369), the Young Scientists of Taishan Scholarship (Grant No. tsqn202408168), the Higher Educational Youth Innovation Science and Technology Program of Shandong Province (2023KJ202). A portion of work was supported by Zhejiang Provincial Natural Science Foundation of China (Grant No. LZ25A040003) and Shandong Provincial Natural Science Foundation (Grant No. ZR2023QA057). R.S. acknowledges the financial support provided by the Ministry of Science and Technology in Taiwan under Project Numbers NSTC-114-2124-M-001-009, NSTC-113-2112-M-001-045-MY3, and Academia Sinica for the budget of AS-iMATE-115-11. R.K.U. would like to acknowledge the IITR for the Faculty Initiation Grant (No. FIG-101068).

C. Zhang, J. J. Feng, and X. Tang contributed equally to this work.